\documentclass[sigconf, nonacm]{acmart}
\newcommand\vldbdoi{XX.XX/XXX.XX}
\newcommand\vldbpages{XXX-XXX}
\newcommand\vldbvolume{XX}
\newcommand\vldbissue{X}
\newcommand\vldbyear{2026}
\newcommand\vldbauthors{Wenkai Dong, Ruyu Li, Sairam Gurajada, Yifan Wang}
\newcommand\vldbtitle{\shorttitle}
\newcommand\vldbavailabilityurl{https://github.com/dongw-netize/NL2Pipe-Compiling-Natural-Language-Questions-into-Semantic-Operator-Pipelines}
\newcommand\vldbpagestyle{plain}

\usepackage{graphicx}
\usepackage{booktabs}
\usepackage{amsfonts}
\usepackage{xcolor}
\usepackage{multirow}
\usepackage[most]{tcolorbox}
\usepackage{dblfloatfix}
\usepackage{microtype}
\usepackage{comment}

\newcommand{\ablate}[1]{\nobreakdash\textendash\hskip0pt #1}

\def\ours{\textsc{NL2Pipe}}

\title{Bridge the Last-Mile Gap to Semantic Analytics: Compiling Natural-Language Queries into Semantic Operator Pipelines}

\author{Wenkai Dong}
\affiliation{%
  \institution{University of Hawaii at Manoa}
}
\email{dongw@hawaii.edu}

\author{Ruyu Li}
\affiliation{%
  \institution{University of Hawaii at Manoa}
}
\email{liruyuluck@gmail.com}

\author{Sairam Gurajada}
\authornote{Work done outside LinkedIn.}
\affiliation{%
  \institution{LinkedIn}
}
\email{sgurajada@linkedin.com}

\author{Yifan Wang}
\affiliation{%
  \institution{University of Hawaii at Manoa}
}
\email{yifanw@hawaii.edu}

\begin{document}

\begin{abstract}
Automated AI workflows increasingly rely on natural-language reasoning over heterogeneous data, but they still lack a practical way to execute such reasoning through optimized semantic data systems. Recent semantic operator systems, such as Palimpzest and LOTUS, expose declarative operators for filtering, joining, mapping, and aggregating over tables, text, images, and other data sources using natural-language predicates. However, these systems require users or workflow developers to manually choose operators, order them, write predicates, and adapt the resulting pipeline to backend-specific APIs. This manual process is difficult for non-experts, brittle across backends, and infeasible for automated workflows where queries and data contexts vary at runtime.

We present \ours{}, a middleware system that compiles natural-language questions into executable semantic operator pipelines. \ours{} treats this task as a three-phase compilation problem. First, a Query-Data Linker grounds question entities against the actual data and discovers implicit bridge entities needed to connect tables, text, images, and other sources. Second, a Semantic Planner produces a backend-agnostic action plan consisting of semantic operators and natural-language predicates. Third, a Code Generator translates the plan into executable code for a target backend using an auto-generated reference document that captures operator signatures, example pipelines, and backend constraints. This design separates data-aware reasoning from backend-specific code generation, enabling the same planning logic to support multiple semantic operator systems.

The evaluation results show that \ours{} substantially improves pipeline quality on complex cross-source workloads. Comparing to baselines, \ours{} achieves higher pipeline quality (e.g., up to 60\% higher F1 score) while maintaining bounded cost and competitive latency for building the pipelines. These results demonstrate that automatic compilation from natural language to semantic operator pipelines is both practical and effective for bringing semantic analytics to non-expert users and automated AI workflows.

\end{abstract}

\maketitle

\pagestyle{\vldbpagestyle}
\begingroup\small\noindent\raggedright\textbf{PVLDB Reference Format:}\\
\vldbauthors. \vldbtitle. PVLDB, \vldbvolume(\vldbissue): \vldbpages, \vldbyear.\\
\href{https://doi.org/\vldbdoi}{doi:\vldbdoi}
\endgroup
\begingroup
\renewcommand\thefootnote{}\footnote{\noindent
This work is licensed under the Creative Commons BY-NC-ND 4.0 International License. Visit \url{https://creativecommons.org/licenses/by-nc-nd/4.0/} to view a copy of this license. For any use beyond those covered by this license, obtain permission by emailing \href{mailto:info@vldb.org}{info@vldb.org}. Copyright is held by the owner/author(s). Publication rights licensed to the VLDB Endowment. \\
\raggedright Proceedings of the VLDB Endowment, Vol. \vldbvolume, No. \vldbissue\ %
ISSN 2150-8097. \\
\href{https://doi.org/\vldbdoi}{doi:\vldbdoi} \\
}\addtocounter{footnote}{-1}\endgroup

\ifdefempty{\vldbavailabilityurl}{}{
\vspace{.3cm}
\begingroup\small\noindent\raggedright\textbf{PVLDB Artifact Availability:}\\
The source code, data, and/or other artifacts have been made available at \url{\vldbavailabilityurl}.
\endgroup
}

\section{Introduction}
\label{sec:intro}

Semantic operator systems~\cite{patel2025lotus, liu2025palimpzest, zhu2025nirvana, wang2025unify, russo2025abacus, dai2024uqequeryengineunstructured} have emerged in recent years which integrate semantic processing functionality into data management systems and frameworks. With a series of builtin \emph{semantic operators} implemented on top of Large Language Model (LLM), like semantic filter (\texttt{sem\_filter}), semantic join (\texttt{sem\_join}), semantic map (\texttt{sem\_map}), semantic aggregation (\texttt{sem\_agg}) and so on, these systems fuse the semantic processing capability into the original data systems like Python Pandas Dataframe or relational databases. 
These operators accept natural language predicates and conduct corresponding operations based on the semantic predicates. Additionally, they are often backed by built-in optimizers to select best models, prompts, and execution plans automatically. 
Users are able to programmatically call these operators to process natural language (NL) queries without external tools. For example, a semantic filter can be called with predicate "the course requires a lot of math" to filter all courses that need solid math background. Such semantic operator systems provide a new paradigm for semantic analytics: neither text-to-SQL + traditional database, nor relying on tools outside the databases, but an advanced in-database solution to process semantics natively.     

\begin{figure*}[tbp]
    \centering
    \includegraphics[width=\textwidth]{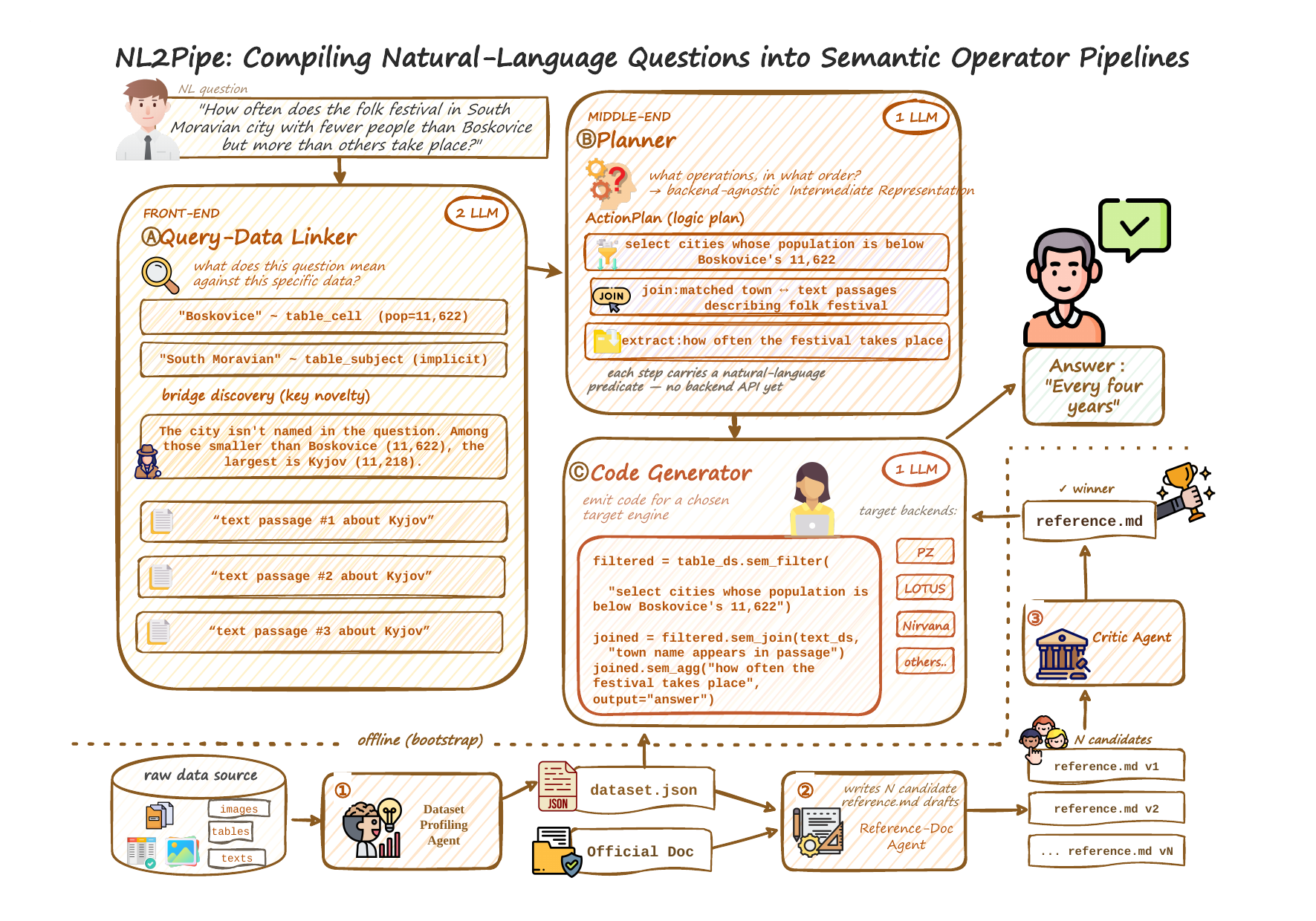}
    \caption{End-to-end architecture of \ours{}, illustrated on the
running example \emph{``How often does the folk festival in the
South Moravian city with fewer people than Boskovice but more
than others take place?''}.
The \textbf{front-end Query Linker} (Phase~A, two LLM calls)
grounds question entities against the data
(\emph{``Boskovice''} maps to a table cell, population~11{,}622)
and discovers bridge entities not named in the question---
ranking by population just below Boskovice yields
\emph{``Kyjov''}, which is verified against text passages
mentioning the folk festival. The grounded entities and matched
passages found by the bridges are then handed to the
\textbf{Semantic Planner} (Phase~B, one LLM call), which emits
a backend-agnostic \emph{ActionPlan}---a sequence of
logical operations with
natural-language predicates. The \textbf{Code Generator}
(Phase~C, one LLM call) compiles the plan into executable code
for a chosen target engine (Palimpzest, LOTUS, Nirvana, etc). Executing the code returns the final answer (\emph{``every four
years''}).
The \textbf{offline bootstrap} (bottom) is a one-time setup per
(backend,~dataset) pair: a Dataset Profiling Agent (A) derives
the dataset summary from raw files, a Reference-Doc Agent
(B) generates $N$ candidate reference documents, and a Critic
Agent (C) selects the winning document used as Phase~C's system
prompt.}

    \label{fig:overview}
\end{figure*}

For complex multi-hop queries, a pipeline consisting of multiple 
operators has to be built. For instance, as shown in Figure~\ref{fig:overview}, consider answering 
``\textit{How often does the folk festival in the South Moravian city 
with fewer people than Boskovice but more than others take place?}'' 
given a table of South Moravian towns with their populations 
(the table source) and a collection of Wikipedia passages 
describing each town (the text source). Answering this requires 
a precise sequence of semantic operators: first, a semantic filter 
over the table to retain rows whose population is strictly less than 
Boskovice's (11{,}622); then, a cross-source semantic join that 
links each surviving row to the passage describing that town and 
locates the single row (target town) whose population is the largest among these candidates; finally, a semantic aggregation or semantic map that extracts the festival's frequency from the matched passage for the target town. Each operator is 
parameterized by a natural-language predicate derived from the 
corresponding task, like \texttt{sem\_agg} uses predicate "how often the festival takes place". 

The challenge is that no existing system can automatically and accurately compile the question into this operator sequence. The user must manually determine which operators to invoke, in what order, with what predicates, and over which data sources. These decisions require understanding not only the question and data semantics, but also the data structure behind the operators and datasets. 
For instance, the descriptive phrase ``fewer people than Boskovice 
but more than others'' must first be resolved against the table 
through a non-trivial ranking inference, which takes the row with the 
\emph{maximum} population among those strictly smaller than 
Boskovice, to identify the bridge entity (Kyjov, population 
11{,}218) that is an anchor to link cross-source data (which are table and text in this example). Specifically, bridge entity is an entity that never appears in the original question, but has to be found as the join key to link different data sources. 
In this example, the name "Kyjov" is not mentioned in the question, but it must serve as the join key linking the filtered table row to the text passage that records the festival's quadrennial cadence. Getting any of these steps wrong leads to missed cross-source connections, redundant processing, or incorrect answers.
For complex queries like multi-hop queries over multi-modal data, the pipelines are usually complicated and making those decisions are time-consuming and hard to be optimal if relying on human.

A bigger challenge is raised since the required operator pipeline is \emph{query-dependent}, i.e., the pipelines to solve each query are possibly different. Some queries only need filtering on table, while some needs cross-source semantic join between different modalities and semantic aggregation on multiple records, etc. Even if two queries have the same form like "Which $X$ has the highest $Y$", their pipelines could be different due to other conditions like data sources involved. Such a query-and-pipeline dynamics makes manually writing pipelines for each query not scalable.  
Additionally, even the decisions for a pipeline have all been made, generating an executable pipeline is still non-trivial: with different semantic operator systems (called \emph{backend systems} in this paper), the APIs and pipeline organizations are different, which requires fundamentally different code, with distinct execution models, column reference syntax, and result extraction patterns. This introduce additional complexity to the pipeline building. 
Furthermore, if people want to use these semantic operator systems to facilitate any automated workflow where queries arrive at runtime (like multi-agent systems which heavily rely on semantic processing for NL queries), manually building the operator pipeline for each different query in the workflow is impossible.   

To address the aforementioned challenges, save human effort in building semantic operator pipelines, and better utilize the operator systems in automated workflows, we propose \textbf{\ours{}}, a middleware that closes the gaps by automatically and accurately compiling natural-language queries into executable semantic operator pipelines. The key insight is to treat the mapping from query to operators as a \emph{compilation} problem with three tasks: understanding the data and query intent, deciding what operations to perform, and generating engine-specific code. Accordingly, \ours{} operates in three phases, illustrated in Figure~\ref{fig:overview}. In
\emph{Phase~A} (Query-Data Linker), the system extracts key
entities from the query, locates them in corresponding data
sources
, and discovers \emph{bridge entities} across different data sources,
i.e., values absent from the query but needed to connect data
sources (e.g., the city name ``Kyjov'' in
Figure~\ref{fig:overview}, which links the population table to
the text passages describing each town's festival). \emph{Phase~B} (Semantic Planner)
produces a backend-agnostic plan: an ordered sequence of query
execution steps (e.g., filter, join, and extract) with
natural-language instructions, informed by the Linker's entity
groundings and verified bridges. \emph{Phase~C} (Code Generator)
translates this plan into executable code for a target backend,
guided by an auto-generated reference document that includes API
signatures, golden examples of operator pipelines, and backend-specific constraints.
The generated code is then executed by the target engine and the
result is returned to the user. To our knowledge, \ours{} is one
of the first middleware systems that bridges the emerging
semantic operator systems with non-expert users and automated workflows.
We evaluate \ours{} on several datasets of multi-modal and
multi-hop question-answering and data analytics tasks, and show
that the pipelines compiled by \ours{} produce significant gains
over general-purposed, fixed-structured, and advanced code-generation
baselines on cross-source workloads. \ours{} reaches up to 63\% higher result accuracy than naive LLM-generated pipelines in the evaluation. 

\ours{} is not a Text-to-SQL system. The two problems differ in their input assumptions, target programs, execution engines, and reasoning scope. Text-to-SQL assumes that the data is stored in a relational database and translates a natural-language question into a standard SQL query. Its main challenge is schema and value grounding: mapping phrases in the question to tables, columns, and cells, then expressing the computation using relational operators. In contrast, \ours{} assumes a heterogeneous data context that may include tables, texts, images, or other modalities, and compiles the question into a pipeline of semantic operators executed by systems like Palimpzest. The generated program is not SQL but an ordered composition of semantic operators whose predicates are natural-language instructions interpreted by LLM. As a result, \ours{} can express operations that SQL cannot naturally capture, such as joining a table row to a passage or image through semantic content and extracting answers from unstructured evidence. Thus, Text-to-SQL solves relational query translation, whereas \ours{} solves semantic pipeline compilation for multi-modal and automated AI workflow settings.

\medskip

Our contributions are as follows:
\begin{itemize}
\item We identify the gap that non-expert users / automated workflows lack effective automated solutions to construct semantic operator pipelines, and propose one of the first middleware to fix the gap by accurately building the pipelines without domain expertise or human intervention.

\item We propose a novel three-phase, reference-document-guided workflow, consisting of data-aware query analysis, backend-agnostic planning, and backend-specific code generation, with optimizations like bridge entity identification and reference document quality enhancement. 

\item We conduct extensive evaluation, showing that our method significantly outperforms general code generation approaches including Codex in terms of pipeline quality and monetary cost while remaining portable across different backends and competitive efficiency. 
\end{itemize}

\section{Related Work}
\label{sec:related}
\subsection{LLM-Based Automated Workflows}
LLM-powered automated workflows increasingly handle complex tasks by
decomposing them into sub-steps executed through tool invocations and
LLM calls. These range from fixed, developer-authored pipelines to fully
autonomous agents that plan their own steps at runtime. Agentic systems
are a prominent instance of this paradigm: representative architectures
include ReAct~\cite{yao2023react}, which interleaves reasoning traces with
actions; HuggingGPT~\cite{shen2024hugginggpt}, which routes sub-tasks to
specialist models; and MetaGPT~\cite{hong2024metagpt}, which assigns
distinct roles to collaborating agents. More broadly, such workflows have
been applied to code generation~\cite{chen2021codex},
text-to-SQL~\cite{pourreza2024dinsql}, and scientific
research~\cite{boiko2023coscientist}.
A shared limitation across these systems is that each reasoning step executes as an independent LLM call with no cross-step optimization. Two strategies dominate: single-pass code generation, which lacks the ability to adapt operator choices per query, and rigid tool-calling sequences, which cannot handle the diversity of data modalities and reasoning patterns encountered in practice. AgenticData~\cite{sun2025agenticdata} 
builds a semantic data analytic system from scratch, implementing everything from storage to planner. Particularly, they implement their own semantic operators rather than using any existing systems. Unlike it, \ours{} acts as a middleware that can adopt to any existing semantic operator systems, including the semantic operators of AgenticData.    

\subsection{Multi-Modal Question Answering}
A large body of work answers multi-hop, multi-modal questions
\emph{directly}. For multi-hop text QA, the dominant method
is retrieve-then-read: iteratively retrieve supporting passages
and feed them to a reader that chains the evidence into an
answer~\cite{karpukhin2020dpr,xiong2021mdr,trivedi2023ircot}. For questions over tables and text, methods
either linearize the table into the reader's input or train a
hybrid reasoner that operates over cells and passages jointly,
sometimes with a numerical-reasoning module for aggregation and
arithmetic~\cite{chen2020hybridqa,kumar-etal-2023-multi,zhu2021tatqa,zhou2022unirpg}. For questions that further span images,
a modality-selection or decomposition step routes each sub-question
to a text, table, or vision module and recombines the partial
answers~\cite{talmor2021multimodalqa,hannan2020manymodalqa,zhang2023moqagpt}.
These methods are effective on semantic query processing in their target tasks, but they are
end-to-end solutions tuned to one benchmark's structure, and they
usually require significant engineering effort to be seamlessly integrated into existing data systems for semantic data operation, which limits their utilization in end-to-end data analytics.
This gap motivates expressing such queries as native \emph{semantic
operators} in a data system, which we review next: doing so
turns an ad-hoc QA model into optimizable, reusable in-database
operations. 

\subsection{Semantic Operator Systems}
Unlike standalone QA solutions, semantic operator systems provide declarative operators implemented on top of LLM, like LLM-powered semantic filter, map, join, and aggregate, which can understand natural language semantics and operate/analyze data based on natural language predicates.  
The operators may additionally pair with optimizers that search over models, prompts, and execution orders. Palimpzest~\cite{liu2025palimpzest} pioneered this paradigm, exploring cost--quality tradeoffs across execution plans. Subsequent systems have extended it along different axes: Abacus~\cite{russo2025abacus} adds cost-based optimization using validation examples; LOTUS~\cite{patel2025lotus} provides formal accuracy guarantees through model cascades; Nirvana~\cite{zhu2025nirvana} supports multi-modal analytics with per-operator backend selection. Related paradigms include DocETL~\cite{shankar2024docetl} for document processing, CAESURA~\cite{urban2024caesura} for multi-modal SQL, SUQL~\cite{liu2024suql} and TAG~\cite{biswal2024tag} for hybrid text-SQL queries, and SemBench~\cite{lao2025sembench} for benchmarking.
Despite their optimization capabilities, all these systems require users to manually compose operator pipelines, i.e., selecting operators, ordering them, and writing natural-language predicates for each. Unify~\cite{wang2025unify} takes a step toward automation by generating query plans directly from natural language, but it targets a single backend and does not handle multi-modal data or cross-source entity linking. \ours{} addresses the remaining gap: it automates the full compilation from question to executable pipeline, discovers bridge entities needed for cross-source reasoning, and targets multiple backends through a single planning framework.

\section{Method}
\label{sec:method}

\subsection{Problem Formulation}
\label{sec:method:problem}

Given a natural-language question $q$ and a heterogeneous data
context which may include a relational table, a collection of
text passages, a collection of images, any other data modalities, or any combination of these, the task is building a semantic operator pipeline which will return an answer to $q$ after being executed, in other words, compiling the query $q$ into such a pipeline.  

We formulate a \emph{semantic operator
pipeline} as such: an ordered composition of declarative semantic operators whose
predicates are natural-language strings and whose semantics are
realized by LLM calls at execution time. Recent semantic operator
systems~\cite{liu2025palimpzest,patel2025lotus,zhu2025nirvana}
expose a variety of such operators, among which the most widely
used are a filter $\sigma_{\phi}$ that retains items satisfying
natural-language predicate $\phi$, a join $\bowtie_{\phi}$ that
pairs items from two collections under predicate $\phi$, and an
aggregation $\gamma_{\phi}$ that reduces a collection of data to a
single result under $\phi$. Additionally, per-row semantic mapping, ranking, and other
specialized operators are also commonly available. 

Based on such operators, given the question $q$, data context $\mathcal{D}$, and a target backend $B$, 
 an operator pipeline $P_B$ is a finite ordered sequence of the operators that is specifically executable on $B$. Executing the pipeline on backend $B$ within the data context $\mathcal{D}$ yields the answer to question $q$.
Particularly, the question and data context are
backend-agnostic while the produced pipeline is backend-specific. To make this mapping ($q, \mathcal{D}, B \rightarrow P_B$) more accurately, we split it into two stages: (1) Intermediate plan generation ($q, \mathcal{D} \rightarrow I$) that generates a backend-agnostic step-by-step plan $I$ about how to answer $q$ based on $\mathcal{D}$, as shown in Figure~\ref{fig:overview} Query-Data Linker and Planner phases, and (2) Plan-to-executable-pipeline mapping ($I, B \rightarrow P_B$), where the backend specification (like operator APIs) is introduced to translate the intermediate plan into executable backend-specific pipeline program, shown as the Code Generator phase in Figure~\ref{fig:overview}.

\subsection{Three-Phase Compilation}
\label{sec:method:overview}

\ours{} factors the compilation process into three phases, each includes a small number of LLM calls and pivoted by an 
intermediate representation. Phase~A discovers bridge
entities that link cross-source data for multi-modal scenarios; Based on the linked data items and their modalities, Phase~B emits a backend-agnostic action plan with abstract data operations and corresponding predicates; Phase~C translates the plan into executable
code for a target backend. Phases~A and~B are fully
backend-agnostic, where the working logic is not impacted by 
the execution backend, and only Phase~C is backend-specific. Comparing to general code generation that produces final code in only one step, this multi-phase approach decouples query processing logic and backend-specific specification, guaranteeing high compatibility (easy switch) between different backends.  

Figure~\ref{fig:overview} illustrates the pipeline on our
running example: \emph{``How often does the folk festival in the
South Moravian city with fewer people than Boskovice but more
than others take place?''}, paired with a table of South
Moravian cities and their populations and a collection of
Wikipedia passages describing each town. Answering this question
requires cross-source reasoning because any single source (table or passages) does not include full information required by the complete reasoning path: the table includes city name and population but not the festival information while the text includes festival information but no population.  
Therefore, finding a pivot entity to link different sources is critical, for which the central difficulty is that 
such entities are often \emph{implicit} (hidden) in the question. 
Like in our running example, the target city ``Kyjov'' is never
named in the question, instead, it must be resolved against the data. And only when we identify it, we can join the information between different data sources (corresponding passages and the table in the example) to enable a complete cross-source reasoning path.  
Such values that must be resolved against the data to allow proper data source joining is called \emph{bridge entities}: they are often absent in the query but necessary to bind operators (like \texttt{sem\_join}) across data sources. Bridge entities guarantee accurate collection of complete key information.   


\subsection{Phase~A: Query-Data Linker}
\label{sec:method:phaseA}

Phase~A prepares the evidence that Phase~B needs to select proper operators and write
precise operator predicates. It performs up to two LLM calls:
one used by \emph{entity grounder} that jointly grounds question entities
and identifies bridge entities, and another optional call used by \emph{bridge verifier} to double check the bridge entities truly exist in  the linked data sources.

\paragraph{Entity grounding and bridge discovery.}
The entity grounder first identifies \emph{question entities} in the question that explicitly exist in the question and may be part of the operator predicates, e.g., noun phrases and numeric constraints. Each candidate entity/phrase will be assigned one of three labels.
(1) A phrase is \emph{matched} if it appears as a value in some
data source (e.g., a cell in the table, or a span in some text
passages or image captions). Matched phrases
will form column-aware filter predicates in Phase~B.
(2) A phrase is \emph{topical} if it describes the subject of a
source rather than appearing as content within it.  Topical
phrases are prohibited in filter predicates, since filtering on them would
eliminate all data items as they never exist in the data content. In the running example, ``South Moravian'' is
topical: it matches the title of the table instead of any cell in the table. Therefore, although such phrases are certainly related to the data sources, they should not be used as keywords/constants in predicates. 
(3) A phrase is \emph{absent} if it is neither matched nor topical
and must therefore be resolved through a different source. 
Particularly, if the data context includes tables, the matched/topical/absent will be labeled only based on the tables without checking text, image or other sources, to avoid LLM overthinking. 
In the running example, ``folk festival'' is absent because there is a table and the table contains no festival information. This phrase has to be
resolved against the text passages other than table.

\begin{figure*}[tbp]
    \centering
    \includegraphics[width=\textwidth]{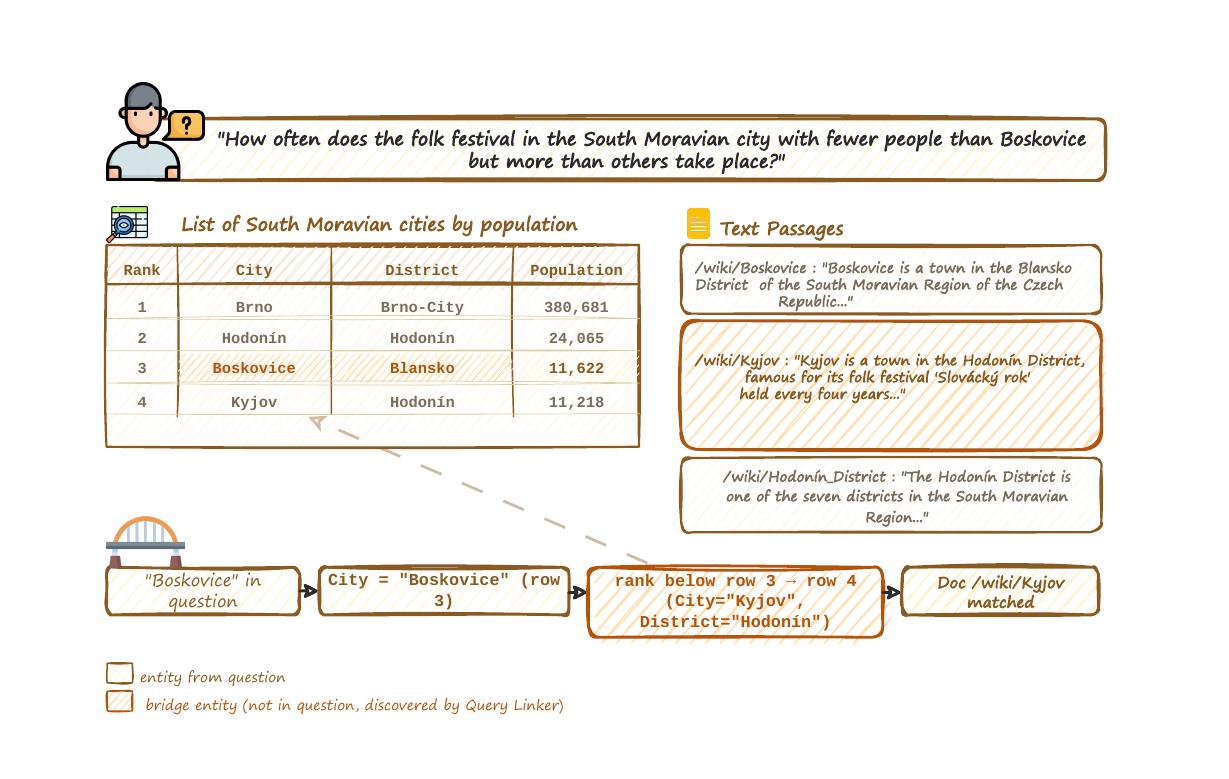}
    \vspace{-12pt}
    \caption{Bridge entity discovery on a HybridQA example. Top: the table and text passages seen by the system. The yellow cell marks the entity 
grounded from the question ("Boskovice" maps to the City of row 3); amber cells mark bridge entities discovered by the Query Linker (City = 
Kyjov, District = Hodonín) — absent from the question but needed to retrieve the relevant passage. Bottom: the step flow shows how "Boskovice"
 anchors row 3, the rank-below row's cells become bridge entities, and bridge "Kyjov" identifies the candidate text passages. The grounded 
entities and matched passages are then handed to the downstream Planner (Phase B), which generates the operator pipeline that ultimately 
extracts the answer.}
\label{fig:bridge}
\end{figure*}

The grounder simultaneously identifies bridge entities by
inspecting the data sources for values that could plausibly connect two
sources. 
Figure~\ref{fig:bridge} zooms in this mechanism for the
running example: the ranking constraint ``fewer people than Boskovice but more than others'' grounds row~4, then its
cell \texttt{City}=Kyjov
become a bridge entity, by which the text passages related to ``Kyjov'' are identified for being joined with the table later. By such bridge entities, different sources are linked to provide complete information.  
Bridge entities usually start from the tables to other sources like text or image collection. 
When no row is directly pinned down, the grounder will identify a column that could serve as
the natural join key with another source, yielding every entry
of that column as a bridge entity from the table to the other source.

After identifying all bridge entities, an additional LLM call could be conducted to verify whether they truly appear in the other source, to reduce the risk of filtering based on non-existing targets. We preprocess all data items offline to generate short previews for each of them, and the verification will let LLM check the existence of the entities against those previews. Non-existing entities will be discarded before sending to Phase~B. 


\subsection{Phase~B: Semantic Planning}
\label{sec:method:phaseB}

Phase~B compiles the question $q$, dataset schema and the Phase~A evidence into a backend-agnostic plan. The evidence consists of three forms of information: matched phrases with
their locations in corresponding data sources, topical phrases which describe sources rather than appear in them, and verified bridges with
their cardinality and supporting document identifiers. Phase~B Planner will consume them to construct a
backend-agnostic step-by-step plan to answer the query, which we call the
\emph{action plan}.

An action plan is a finite sequence of steps, where each step
is a triple $(o_i, \mathrm{src}_i, \phi_i)$: $o_i \in
\{\sigma, \bowtie, \gamma\}$ is the operator type,
$\mathrm{src}_i$ identifies the source(s) the operator applies to, and
$\phi_i$ is a natural-language predicate attached to the operator. 
Our current implementation supports \texttt{sem\_filter} ($\sigma$), \texttt{sem\_join} ($\bowtie$) and \texttt{sem\_agg} ($\gamma$) which are three most commonly used semantic operators that can handle most queries. 
An action plan does not reference any backend API, instead, representation of each step is a loosely structured natural language description for the operation and predicate, as shown in Figure~\ref{fig:overview} Phase~B.  

Generating this action plan (i.e., the planning) uses a single LLM call over four inputs: the question, dataset schema, the Phase~A output, and
a short list of backend-agnostic planning hints that encode
failure modes empirically observed across all backends (for
example, text filters should use broad entity-based predicates
joined by disjunction rather than narrow conjunctive
conditions, which downstream operators handle poorly). Phase~A's
three forms of evidence are wired into the plan in distinct
ways: matched phrases become precise column-referencing filter
predicates, topical phrases are excluded from all filter
conditions to avoid eliminating all rows, and verified bridges
become either filter anchors or join keys depending on their
cardinality.

For the running example, Phase~B emits a three-step plan:
filter the table to retain cities whose population is below
Boskovice's 11{,}622, join the matched town with the text
passages describing the folk festival, and aggregate over the
joined records to extract how often the festival takes place.

The separation of planning from code generation is deliberate:
under a single combined prompt, the model frequently drops
join steps or produces type errors when forced to juggle
abstract plan logic and concrete syntax simultaneously.
Treating the action plan as an explicit intermediate
representation also makes the system portable across backends,
since only the final translation step changes.

\subsection{Phase~C: Code Generation}
\label{sec:method:phaseC}

Phase~C translates the action plan into executable code for a
target backend. The translation is a single LLM call with two
inputs: the action plan, and an auto-generated \emph{reference
document} placed in the system prompt.

The reference document is pre-generated offline and organizes everything Phase~C needs into
three layers, illustrated  in
Figure~\ref{fig:refdoc} which uses that of Palimpzest as an example.
\textbf{Layer~1} lists the backend's operator signatures, so
the model knows what functions are available and the parameters for each function.
\textbf{Layer~2} provides one runnable golden pipeline per
\emph{question type}, where a question type is defined by the
data source modalities required to answer it, e.g., TableQ can be 
answered from a table alone, TextQ from text passages, ImageQ
from images, and Compose(TextQ,\allowbreak{} TableQ) requires cross-source
reasoning between text and table. Each example shows a complete
end-to-end pipeline for that type, so the model can pattern-match
its plan to the closest precedent and adapt the predicates rather
than synthesizing the API calls from scratch.
\textbf{Layer~3} appends runtime constraints, i.e., format
requirements, common pitfalls, and engine-specific
invariants, which the generated code must satisfy. 

For our running example, Phase~C receives the three-step action
plan from Phase~B (filter $\rightarrow$ join $\rightarrow$
aggregate) and consults the Compose(TextQ,\allowbreak{} TableQ) entry in
Layer~2 of the reference document. Then the model adapts the Phase~B generated template by substituting the plan's natural-language predicates
into the corresponding operator API calls and emits the final backend-specific program. The generated program, when executed,
invokes the backend's implementation of the semantic operators to operate data and generate result.  
The execution time and LLM cost are then managed by the backends themselves and lie outside our scope.

\paragraph{Auto-generated reference documents.}
The reference document (Figure~\ref{fig:refdoc}) is produced
offline, once per (backend,~dataset) pair, by a bootstrap
pipeline (bottom of Figure~\ref{fig:overview}). So it does not introduce runtime overhead and cost.  
Specifically, a Dataset
Profiling agent inspects the raw dataset to produce a
structured profile of its data sources. A Reference-Doc
agent then generates multiple candidate documents based on the
backend's official API documentation and the dataset profile. By this point, each candidate document is composed of operator signatures (Layer~1) and golden pipeline examples (Layer~2). Finally, a Critic agent scores
all candidates against the backend's own example code (officially provided in their documentation or repository) on
criteria such as syntactic correctness, pipeline completeness,
and coverage of all question types, and selects the
highest-scoring candidate. Finally, runtime constraints (Layer~3) like format
requirements, common pitfalls, and engine-specific invariants are
appended to the selected document as-is. 

\begin{figure}[tbp]
\centering
\definecolor{rdBorder}{HTML}{D6B656}
\definecolor{rdFill}{HTML}{FFF2CC}
\definecolor{rdAccent}{HTML}{9C7A2E}

\begin{tcolorbox}[
  enhanced,
  colback=white,
  colframe=rdBorder,
  boxrule=0.6pt,
  arc=2pt,
  left=4pt, right=4pt, top=3pt, bottom=3pt,
  fonttitle=\footnotesize\bfseries,
  coltitle=black!85,
  colbacktitle=rdFill,
  title={Reference Doc\hfill\normalfont\tiny\itshape Palimpzest, MMQA},
]

{\footnotesize\textbf{\textcolor{rdAccent}{Layer 1: API Signatures}}}\\[1pt]
{\ttfamily\tiny%
dataset.sem\_filter(filter, depends\_on)\\
dataset.sem\_join(other, condition, depends\_on)\\
dataset.sem\_agg(col, agg, depends\_on)
}

\vspace{3pt}\textcolor{rdBorder!60}{\rule{\linewidth}{0.3pt}}\vspace{1pt}

{\footnotesize\textbf{\textcolor{rdAccent}{Layer 2: Golden Pipeline (TableQ)}}}\\[1pt]
{\ttfamily\tiny%
table\_ds = pz.MemoryDataset(id="t", vals=table\_rows)\\
result = table\_ds.sem\_agg(col=\{...\}, agg=f"\{question\} \{INSTR\}")
}

\vspace{3pt}\textcolor{rdBorder!60}{\rule{\linewidth}{0.3pt}}\vspace{1pt}

{\footnotesize\textbf{\textcolor{rdAccent}{Layer 3: Runtime Constraints}}}\\[1pt]
{\tiny%
$\bullet$~Call \texttt{\_inject\_schema} after every \texttt{MemoryDataset}.\\
$\bullet$~\dots
}

\end{tcolorbox}
\caption{Overview of an example auto-generated Reference Doc
for Palimpzest, MMQA. Three layers: operator signatures,
one runnable pipeline per question type, and 
runtime constraints. Full document is $\sim$250 lines.}
\label{fig:refdoc}
\end{figure}

\subsection{Cost Summary}
\label{sec:method:cost}

The per-query cost of \ours{} is bounded: three LLM calls
for single-source query (the Linker's entity-grounding
step, plus one for the Planner and one for Code Generator) and four for
composed questions where the Linker may additionally runs bridge
verification. This overhead is tiny comparing to the
internal LLM calls inside the backend during
execution, which typically number in dozens and dominate
end-to-end cost. The offline bootstrap (dataset profiling
and reference-document generation) adds a one-time cost per
(backend,~dataset) pair that is amortized across all
subsequent questions.
Our evaluation reports this online mapping cost (from query until the pipeline built) separately from
backend execution cost, and shows that the additional compilation
overhead buys substantial quality gains on complex workloads.

\section{Evaluation}
\label{sec:eval}

We evaluate \ours{} along four dimensions: (i)~\emph{end-to-end
answer quality} after running the pipelines it produces,
(ii)~\emph{mapping cost} which is the LLM cost spent in building the pipelines, (iii)~\emph{efficiency}, including mapping efficiency for building the pipelines and execution efficiency for running the pipelines, and (iv)~\emph{component contribution} via ablation study.

\subsubsection{Datasets and Backends}
\label{sec:eval:data}
We use five evaluation datasets:
MMQA~\cite{talmor2021multimodalqa},
HotpotQA~\cite{yang2018hotpotqa},
HybridQA~\cite{chen2020hybridqa},
ManyModalQA~\cite{hannan2020manymodalqa},
and TAT-QA~\cite{zhu2021tatqa}.
They span three modalities (table, text, image) and several
reasoning patterns. MMQA and ManyModalQA both involve all three
modalities, but MMQA requires multi-hop reasoning whereas
ManyModalQA is mostly single-hop. HybridQA pairs every question
with a Wikipedia table and linked passages, so all its questions
are inherently cross-source. HotpotQA is text-only and tests
multi-hop reasoning without table structure. TAT-QA adds
numerical reasoning over financial reports with hybrid
table--text contexts. Together, these datasets cover diverse scenarios and complexity, ensuring the conclusion is strong enough. 

Because semantic-operator execution is slow (many LLM calls per
question), we evaluate on a stratified sample of 300 questions
per dataset, drawn proportionally to question types (e.g.,
TableQ/TextQ/ImageQ/Compose for MMQA; bridge/comparison for
HotpotQA). This preserves coverage of fine-grained types
(e.g., MMQA has totally 15 types of queries) while keeping the grid
tractable: the full $5~\text{major\_baselines}\times3\text{ backends}\times5\text{ datasets}$ evaluation already requires $22{,}500$ mapping LLM calls plus far more backend-internal calls, in which case running all of them on the full datasets would take roughly $104$ days. 

We choose three state-of-the-art semantic operator systems as evaluation backends: Palimpzest~\cite{liu2025palimpzest}, LOTUS~\cite{patel2025lotus} and Nirvana~\cite{zhu2025nirvana}. Each question and pipeline will be tested using all these systems. 

 \subsubsection{Baselines}
 \label{sec:eval:baselines}
 We compare \ours{} against five baselines that span the design
 space of LLM-driven pipeline construction.
 
(1) \textbf{Naive} issues a single LLM call that receives the
 backend's official documentation, a dataset description, and
 the question, and directly emits executable code.

(2) \textbf{Hardcoded} is a baseline simulating that a human
engineer hand-writes one pipeline \emph{template} per question type
(e.g., \emph{filter $\rightarrow$ aggregate} for TableQ;
\emph{filter $\rightarrow$ join $\rightarrow$ aggregate} for Compose) and phrases
the natural-language predicate into each operator within the template. In our implementation an LLM performs the predicate composition at query time to simulate human doing it. And because this is just simulation which is actually completed by human in real world, its token cost does not make sense and is not comparable to other methods. We therefore report Hardcoded's answer quality and latency, while omit its mapping cost. And actually the mapping latency is also not comparable to other methods due to the same reason, while its pipeline execution latency is still meaningful. 
Hardcoded estimates the pipeline quality attainable by hand-crafted
templates. 
This baseline proves that pre-defined templates are far from enough to construct effective pipelines for complex workload.  

(3) \textbf{CodeTree}~\cite{li2024codetree} is an agent-guided
 code-search baseline. For each question it proposes up to five
 candidate strategies, generating and executing code for each. If error occurs, additional reflection will inspect and
 propose a repair, which is then re-executed. We allow the
 search to expand to at most three branching levels and to take
 at most ten total propose-execute-repair iterations before
 returning the best candidate seen. Unlike \ours{} that does not need execution and loops during the pipeline construction, CodeTree relies on execution feedback and reflection. We use it to evaluate whether such feedback loops truly have
 advantage over \ours{}'s single-pass compilation.

(4) \textbf{Codex (model)} uses the same prompt as Naive but 
calls specific codex model (\texttt{gpt-5.1-codex-mini}) instead of general-purpose reasoning model used by the other baselines. It serves as a stronger baseline than Naive as the codex model is optimized for code generation. We use this baseline to evaluate
 how well a standalone code-specialized model can handle our task.

(5) \textbf{Codex (agent)} is the standard Codex agent that relies on the same code-optimized model
(\texttt{gpt-5.1-codex-mini}) with a best-of-three harness:
for each question the agent generates three candidate pipelines independently with full tool access, executes each candidate using the backend, and picks the best of the three answers based on the agent's understanding of the query and data. We set it to generate 3 candidates based on grid search from 1 to 3, where at 3 the agent has already needed longer running time than \ours{} but still underperforms \ours{}. So we stop at 3.   
Because each query triggers three full generations and executions, it is markedly slow and costly, so we evaluate it on $100$ stratified sample queries per (backend,~dataset) combination (reported in Table~\ref{tab:codex_3way}) rather than using the full 300 queries.

 In our implementation, for
 pipeline construction, the Codex (model) and Codex (Agent) baselines use \texttt{gpt-5.1-codex-mini}, while all other baselines and \ours{} use \texttt{gpt-5-mini}. The backend engines are always using \texttt{gpt-5-mini} for pipeline execution.  

\subsubsection{Metrics}
\label{sec:eval:metrics}
To measure the pipeline construction quality, each pipeline will be executed using the corresponding backend and get the returned answer. Based on the generated answer, we report three quality metrics, all shown as percentage without explicit "\%":
(1) \textbf{Exact Match} (EM) is the fraction of generated answers exactly matching the gold answers on string level, after standard normalization (lowercasing, punctuation and article removal, and whitespace collapse);
(2) \textbf{token-level F1} is the harmonic mean of unigram
precision and recall against the gold answers; and (3) 
\textbf{LLM-as-judge} (shown as ``LLM'' in tables) is the fraction of generated answers judged semantically
equivalent to the gold answers by an LLM using a fixed six-rule
prompt~\cite{zheng2023judging}. Each quality metric is averaged over all 300 questions. 

We also report average \textbf{mapping cost} in USD per question: the money cost of LLM calls to build each pipeline (which is essentially the process of mapping initial query to the final pipeline). In \ours{}, mapping process is Phases~A--C; in baseline methods, it is the whole process from beginning until the pipeline code is generated, since they do not have offline processing. The cost of backend execution is excluded as it is managed by each backend engine and out of scope of our work. 
Cost of the \textit{Hardcoded} baseline is also excluded, for the reason given in Section~\ref{sec:eval:baselines}.

For efficiency we measure three average latencies per question  (in seconds):
\textbf{mapping latency} $t_m$ (the mapping process),
\textbf{execution latency} $t_e$ (running the generated code using specific backend engine until the result is returned), and
\textbf{end-to-end latency} $t=t_m+t_e$. 
To better present the cost--quality trade-off, we additionally report \textbf{quality-per-unit-cost}, i.e., EM/\$, F1/\$, and LLM/\$ (quality score per milli-USD, where the unit of cost is 0.1 cents), higher is better.

\subsection{Main Results}
\label{sec:eval:main}

Tables~\ref{tab:main_pz}, \ref{tab:main_lotus},
and~\ref{tab:main_nirvana} report end-to-end query processing
quality and mapping cost on Palimpzest, LOTUS, and Nirvana
respectively. 
Due to the long running time of Codex(agent), we only run it on 100 stratified samples of queries over 6 (backend, dataset) combinations, instead of all 300 queries. So we have a separate table (Table~\ref{tab:codex_3way}) for Codex(model), Codex(agent) and \ours{} on those 100 queries. 

\begin{table*}[tbp]
\centering
\small
\setlength{\tabcolsep}{3pt}
\caption{End-to-end results on the \textbf{Palimpzest} backend.
Quality (average EM, F1, LLM) is percentage number without explicit \%; \$ means average mapping cost per
query (in USD). 
Hardcoded's mapping cost is excluded
because the LLM calls in it is purely for simulating human actions, which is not necessary in real cases that manually build the pipelines.   
\textbf{Bold} marks the best quality score
among all pipelines and the lowest mapping cost among pipelines, while \textbf{underline} marks the second lowest mapping cost. }
\label{tab:main_pz}
\resizebox{\textwidth}{!}{%
\begin{tabular}{l cccc cccc cccc cccc cccc}
\toprule
& \multicolumn{4}{c}{\textbf{MMQA}}
& \multicolumn{4}{c}{\textbf{HybridQA}}
& \multicolumn{4}{c}{\textbf{HotpotQA}}
& \multicolumn{4}{c}{\textbf{ManyModalQA}}
& \multicolumn{4}{c}{\textbf{TAT-QA}} \\
\cmidrule(lr){2-5} \cmidrule(lr){6-9} \cmidrule(lr){10-13}
\cmidrule(lr){14-17} \cmidrule(lr){18-21}
\textbf{Pipeline}
& EM & F1 & LLM & \$
& EM & F1 & LLM & \$
& EM & F1 & LLM & \$
& EM & F1 & LLM & \$
& EM & F1 & LLM & \$ \\
\midrule
Naive
& 12.7 & 13.6 & 16.0 & \textbf{0.0052}
&  6.7 &  9.6 & 10.3 & \textbf{0.0054}
& 46.3 & 62.1 & 77.3 & \textbf{0.0045}
& 22.7 & 27.0 & 32.7 & \textbf{0.0049}
& 18.0 & 23.3 & 33.0 & \textbf{0.0053} \\
Hardcoded
& 34.3 & 42.3 & 52.7 & ---
& 39.3 & 47.5 & 51.0 & ---
& 51.7 & 68.9 & 84.3 & ---
& 45.3 & 57.0 & \textbf{71.0} & ---
& 24.0 & 40.5 & 64.7 & --- \\
CodeTree
& 19.7 & 23.7 & 28.7 & 0.0332
& 10.7 & 13.9 & 16.7 & 0.0241
& 33.0 & 43.1 & 57.3 & 0.0283
& 31.7 & 38.7 & 51.0 & 0.0191
& 21.0 & 28.4 & 49.7 & 0.0212 \\
Codex (model)
& 25.3 & 29.7 & 39.0 & \underline{0.0118}
& 22.3 & 27.2 & 30.3 & 0.0164
& 35.3 & 46.9 & 60.3 & \underline{0.0088}
& 36.7 & 45.6 & 61.7 & 0.0105
& 17.3 & 23.5 & 40.0 & 0.0155 \\
\midrule
\ours{} (ours)
& \textbf{51.7} & \textbf{57.1} & \textbf{63.3} & 0.0134
& \textbf{59.3} & \textbf{68.6} & \textbf{73.3} & \underline{0.0156}
& \textbf{64.7} & \textbf{79.3} & \textbf{93.3} & 0.0109
& \textbf{48.7} & \textbf{59.0} & 68.7 & \underline{0.0090}
& \textbf{45.7} & \textbf{55.2} & \textbf{67.7} & \underline{0.0125} \\
\bottomrule
\end{tabular}%
}
\end{table*}

\begin{table*}[tbp]
\centering
\small
\setlength{\tabcolsep}{3pt}
\caption{End-to-end results on the \textbf{LOTUS} backend.
Columns and conventions as in Table~\ref{tab:main_pz}.}
\label{tab:main_lotus}
\resizebox{\textwidth}{!}{%
\begin{tabular}{l cccc cccc cccc cccc cccc}
\toprule
& \multicolumn{4}{c}{\textbf{MMQA}}
& \multicolumn{4}{c}{\textbf{HybridQA}}
& \multicolumn{4}{c}{\textbf{HotpotQA}}
& \multicolumn{4}{c}{\textbf{ManyModalQA}}
& \multicolumn{4}{c}{\textbf{TAT-QA}} \\
\cmidrule(lr){2-5} \cmidrule(lr){6-9} \cmidrule(lr){10-13}
\cmidrule(lr){14-17} \cmidrule(lr){18-21}
\textbf{Pipeline}
& EM & F1 & LLM & \$
& EM & F1 & LLM & \$
& EM & F1 & LLM & \$
& EM & F1 & LLM & \$
& EM & F1 & LLM & \$ \\
\midrule
Naive
& 24.8 & 29.2 & 34.8 & \textbf{0.0059}
& 32.7 & 42.6 & 51.7 & \textbf{0.0057}
& 21.0 & 29.6 & 39.0 & \textbf{0.0055}
& 31.0 & 39.7 & 52.3 & \textbf{0.0055}
& 19.7 & 27.7 & 45.0 & \textbf{0.0055} \\
Hardcoded
&  7.7 & 10.9 & 16.4 & ---
& 20.3 & 30.7 & 47.0 & ---
& 22.7 & 43.7 & \textbf{80.0} & ---
& 23.0 & 28.5 & 40.0 & ---
& 14.0 & 20.0 & 42.7 & --- \\
CodeTree
& \textbf{27.4} & \textbf{36.3} & 47.8 & 0.0316
& \textbf{36.0} & 46.3 & 58.7 & 0.0249
& \textbf{28.7} & 42.4 & 61.7 & 0.0245
& 40.7 & 49.7 & 68.0 & 0.0202
& 29.7 & 40.7 & 61.7 & 0.0245 \\
Codex
& 10.7 & 15.3 & 31.4 & 0.0191
& 11.3 & 20.9 & 45.3 & 0.0195
& 11.3 & 23.5 & 54.0 & 0.0138
& 11.7 & 21.2 & 55.3 & 0.0153
&  4.3 & 13.1 & 52.0 & 0.0127 \\
\midrule
\ours{} (ours)
& 24.4 & 34.7 & \textbf{51.2} & \underline{0.0133}
& 32.3 & \textbf{46.3} & \textbf{66.3} & \underline{0.0177}
& 28.3 & \textbf{45.1} & 74.7 & \underline{0.0112}
& \textbf{43.0} & \textbf{52.6} & \textbf{68.3} & \underline{0.0100}
& \textbf{36.3} & \textbf{47.6} & \textbf{70.7} & \underline{0.0126} \\
\bottomrule
\end{tabular}%
}
\end{table*}

\begin{table*}[tbp]
\centering
\small
\setlength{\tabcolsep}{3pt}
\caption{End-to-end results on the \textbf{Nirvana} backend.
Columns and conventions as in Table~\ref{tab:main_pz}.}
\label{tab:main_nirvana}
\resizebox{\textwidth}{!}{%
\begin{tabular}{l cccc cccc cccc cccc cccc}
\toprule
& \multicolumn{4}{c}{\textbf{MMQA}}
& \multicolumn{4}{c}{\textbf{HybridQA}}
& \multicolumn{4}{c}{\textbf{HotpotQA}}
& \multicolumn{4}{c}{\textbf{ManyModalQA}}
& \multicolumn{4}{c}{\textbf{TAT-QA}} \\
\cmidrule(lr){2-5} \cmidrule(lr){6-9} \cmidrule(lr){10-13}
\cmidrule(lr){14-17} \cmidrule(lr){18-21}
\textbf{Pipeline}
& EM & F1 & LLM & \$
& EM & F1 & LLM & \$
& EM & F1 & LLM & \$
& EM & F1 & LLM & \$
& EM & F1 & LLM & \$ \\
\midrule
Naive
&  1.3 &  2.2 &  2.7 & \textbf{0.0061}
&  8.7 & 10.8 & 11.3 & \textbf{0.0059}
& 16.0 & 23.7 & 33.3 & \textbf{0.0059}
& 36.3 & 44.6 & 57.7 & \textbf{0.0055}
& 20.3 & 27.2 & 40.0 & \textbf{0.0061} \\
Hardcoded
&  0.3 &  0.3 &  0.3 & ---
&  9.7 & 12.1 & 13.7 & ---
& 30.0 & 46.3 & 70.3 & ---
& 24.0 & 27.8 & 32.3 & ---
&  7.0 &  7.5 &  8.7 & --- \\
CodeTree
&  6.3 &  8.7 & 10.3 & 0.0305
& \textbf{31.0} & \textbf{38.8} & \textbf{45.7} & 0.0264
& 30.3 & 40.7 & 54.3 & 0.0214
& 40.3 & 48.1 & \textbf{62.7} & 0.0127
& 24.3 & 33.3 & \textbf{50.7} & 0.0196 \\
Codex
&  7.7 & 13.2 & 25.7 & 0.0153
&  6.7 & 10.5 & 15.7 & \underline{0.0173}
& 15.7 & 25.7 & 49.0 & \underline{0.0102}
& 21.3 & 33.5 & 57.3 & \underline{0.0082}
& 10.0 & 19.3 & 43.3 & \underline{0.0132} \\
\midrule
\ours{} (ours)
& \textbf{27.7} & \textbf{31.1} & \textbf{36.0} & \underline{0.0149}
& 29.3 & 35.4 & 40.0 & 0.0179
& \textbf{57.7} & \textbf{69.0} & \textbf{81.0} & 0.0121
& \textbf{42.7} & \textbf{50.2} & 56.7 & 0.0102
& \textbf{29.3} & \textbf{37.4} & 46.7 & 0.0152 \\
\bottomrule
\end{tabular}%
}
\end{table*}

\textbf{\ours{} achieves the best quality among all methods.} In the full 300-query end-to-end evaluation (Table~\ref{tab:main_pz}, \ref{tab:main_lotus} and \ref{tab:main_nirvana}), the pipeline \ours{} achieves the best quality in most cases with a significantly large advantage over the baselines, especially in complex multi-hop or multi-modal cases where long and cross-source reasoning is needed.  
On Palimpzest, \ours{} perform the highest EM, F1 and LLM-as-judge scores in almost all datasets, where the largest gaps between \ours{} and baselines appear in HybridQA
(20\%, 21.1\% and 22.3\% higher EM, F1 and LLM-as-judge scores than the second-best baseline), 
MMQA (17.4\%, 14.8\% and 10.6\% higher quality than second-best baseline) and TAT-QA (21.7\%, 14.7\%, 3\% higher), which are multi-modal and multi-hop, or require cross-source numerical reasoning.   
In contrast, on text-only HotpotQA and single-hop ManyModalQA, the gaps between \ours{} and baselines are smaller. These prove the effectiveness of our method to build high-quality pipelines, especially for complex workload. 
Similar pattern appears across backends. On LOTUS \ours{} achieves the best quality on HybridQA and TAT-QA, as well as ManyModalQA. On Nirvana our best performance occurs on all datasets except HybridQA.     
Furthermore, as reported in Table~\ref{tab:codex_3way}, \ours{} outperforms the Codex coding agent in most cases with significantly higher quality, lower mapping latency, and 1/30~1/20 cost comparing to Codex agent. This result strongly supports the effectiveness of our method.  

\textbf{\ours{} is cost efficient.} Naive is
the cheapest method but often loses substantial quality, while CodeTree can recover quality only by spending much more on search, and Codex agent takes biggest time and cost due to overcomplex agent runtime and multiple generations for each pipeline. \ours{} is the second-cheapest method in many cases with the highest quality: it achieves a strong overall quality profile
at roughly $\$0.013$ per question (averaging on all datasets and backends), $\sim\!1.8\times$ cheaper than
CodeTree ($\$0.024$), comparable to single-call Codex ($\$0.014$) and $20\sim30$ times cheaper than Codex agent. This is because mapping cost of \ours{} is bounded: the Linker issues at most two LLM calls, followed by one Planner call and one Code
Generator call, regardless of question complexity. CodeTree's
cost grows with reflection rounds and exceeds \$0.03 per query where many candidate strategies fail the verification.

\begin{figure*}[tbp]
    \centering
    \includegraphics[width=\textwidth]{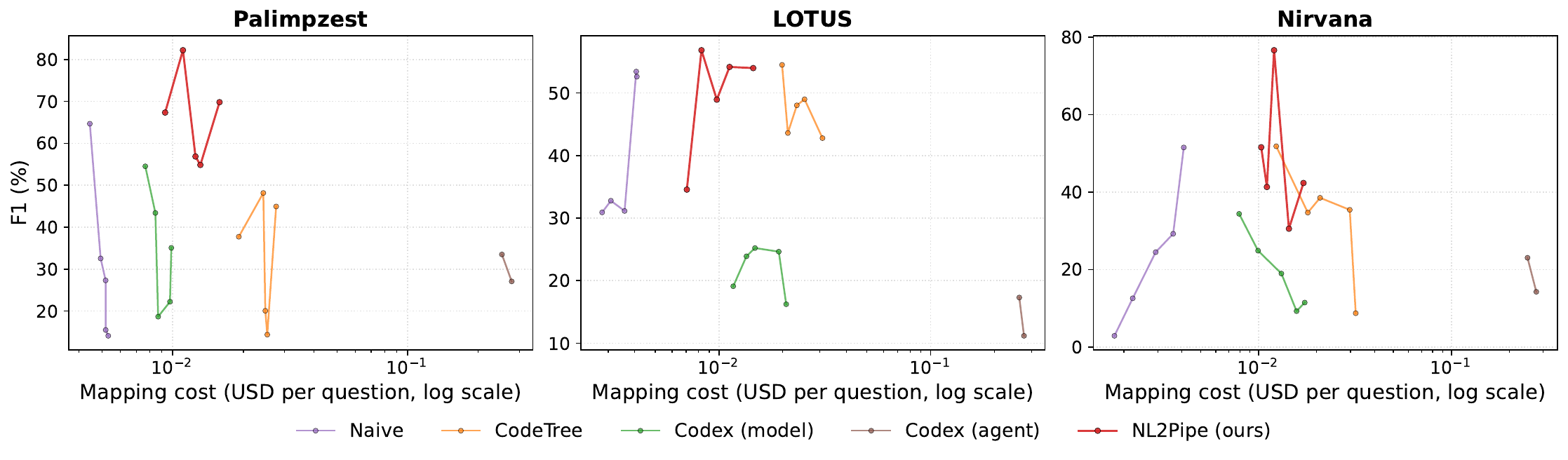}
    \caption{Quality--cost trade-off on the three backends over varying datasets.
Each point is a (pipeline, dataset) combination from
Tables~\ref{tab:main_pz}--\ref{tab:main_nirvana}; same-color
points trace one pipeline across the five datasets. \ours{}
(red) occupies the upper-left portion of the figure in most cases, indicating a better quality-cost trade-off than other methods. Naive is often unavailable due to low quality.}
\label{fig:pareto}
\end{figure*}

Figure~\ref{fig:pareto} visualizes the quality-cost trade-off with F1 and mapping cost, over varying datasets: on all three
backends, \ours{} lies near the upper-left portion, outperforming all baselines except Naive, and Naive is often unavailable due to low quality in most time. This highlights that \ours{} can reach same quality with less cost, or use same cost to get higher quality.   
The offline bootstrap (data profiling and reference doc generation) for \ours{} (Section~\ref{sec:method:phaseC}) adds a one-time cost of \$0.40--\$0.70 per (backend, dataset) pair, which adds only $\$0.001\sim0.002$ to each of the 300 queries and does not impact the conclusions. When more queries come, impact of this cost becomes even less. 

\subsubsection{Failure Mode in Baselines}
We analyze the failure modes for the baselines and discuss how our three-phase architecture solves them.

\textbf{Naive: } A single LLM call must simultaneously decide \emph{what} operators to invoke and \emph{how} to write them according to
the target API. Under this dual burden the generated programs
often make more mistakes, like dropping joins or emitting conjunctive text filters that match no passages. Our Phase~B Planner removes this burden by splitting the two decisions into planning and generation steps, such that in each single step LLM only focuses on one decision, improving accuracy.

\textbf{Hardcoded(Manual Programming): }
Fixed templates are limited and cannot handle many scenarios, like they cannot dynamically look for bridge entities for cross-source operators. Although we use LLM in this baseline to simulate the human programming progress, such programming in real world scenarios is manual and takes much more time than in  simulation. So automated solution is necessary. 

\textbf{CodeTree: } Execution feedback repairs syntactic errors but is hard to handle semantic or logic errors which do not break the pipeline execution but just lead to wrong results.   
CodeTree's reflection loop therefore burns extra
LLM calls (2x more cost per question than \ours{}) on candidates that were doomed from the first generation. In contrast, \ours{} pays more attention to guaranteeing LLM has as complete information as possible at the beginning, such that LLM can see a broader picture about the datasets, backends and semantics hidden in question and data, by preprocessing like identifying query and bridge entities before planning. Our evaluation shows that such beforehand information completeness is more important than reflection and retry for pipeline built with fixed-scope operators.    

\textbf{Codex: }
As a single-LLM call too, Codex (model) achieves better quality than Naive given its code-generation-optimized model, and unavoidably inherits Naive's failure mode, limiting its effectiveness. So it is often outperformed by CodeTree and \ours{}. 
Codex (agent) achieves higher quality than Codex(model) in cost of significantly more latency and cost, and it is still underperforming \ours{}. This is because as a general-purpose coding agent, Codex agent has extra components and advanced architecture that are overcomplex to the pipeline construction task. In such a fixed-scope problem (e.g., limited scope of operators and limited combinations of them), our evaluation shows that Codex agent tends to build unnecessarily complex and long pipelines to solve even the simplest queries. This proves specialized problem needs specialized solutions like \ours{}.    

\begin{table}[tbp]
\centering\scriptsize\setlength{\tabcolsep}{2.5pt}\renewcommand{\arraystretch}{0.95}
\caption{Codex (model) vs.\ Codex (agent) vs.\ \ours{} on $N{=}100$
stratified queries per dataset. EM/F1/LLM-as-judge scores are percentage without \%; \$~is average mapping
cost per query (USD); $t_m$ is average mapping latency per query (s), excluding failed execution runs; \textbf{EM/\$}, \textbf{F1/\$}, \textbf{LLM/\$} are quality
scores per unit mapping cost (milli-USD), higher is better. Datasets abbreviated
\textbf{HyQ}=HybridQA, \textbf{HoQ}=HotpotQA, \textbf{MQ}=MMQA,
\textbf{TQ}=TAT-QA, \textbf{MMQ}=ManyModalQA. \textbf{Bold} marks the best
value for quality and quality-per-cost.}
\label{tab:codex_3way}
\resizebox{\columnwidth}{!}{%
\begin{tabular}{@{}ll l ccc r r ccc@{}}
\toprule
\textbf{Back.} & \textbf{DS} & \textbf{Method} & EM & F1 & LLM & \$ & $t_m$ & EM/\$ & F1/\$ & LLM/\$ \\
\midrule
\multirow{3}{*}{Palimpzest} & \multirow{3}{*}{HyQ} & Codex (model) & 24.0 & 27.9 & 28.0 & 0.0077 & 17.3 & 3.1 & 3.6 & 3.6 \\
 &  & Codex (agent) & 43.0 & 52.9 & 60.0 & 0.2521 & 311.2 & 0.2 & 0.2 & 0.2 \\
 &  & \ours{} & \textbf{57.0} & \textbf{67.3} & \textbf{74.0} & 0.0156 & 111.3 & \textbf{3.7} & \textbf{4.3} & \textbf{4.7} \\
\midrule
\multirow{3}{*}{Palimpzest} & \multirow{3}{*}{MQ} & Codex (model) & 30.0 & 33.7 & 43.0 & 0.0096 & 16.1 & 3.1 & 3.5 & \textbf{4.5} \\
 &  & Codex (agent) & 38.0 & 43.3 & 50.0 & 0.2775 & 143.0 & 0.1 & 0.2 & 0.2 \\
 &  & \ours{} & \textbf{47.0} & \textbf{53.0} & \textbf{59.0} & 0.0137 & 76.8 & \textbf{3.4} & \textbf{3.9} & 4.3 \\
\midrule
\multirow{3}{*}{LOTUS} & \multirow{3}{*}{TQ} & Codex (model) & 7.0 & 14.7 & 51.0 & 0.0132 & 14.8 & 0.5 & 1.1 & 3.9 \\
 &  & Codex (agent) & 7.0 & 17.9 & 66.0 & 0.2758 & 189.9 & 0.0 & 0.1 & 0.2 \\
 &  & \ours{} & \textbf{41.0} & \textbf{52.3} & \textbf{74.0} & 0.0107 & 87.1 & \textbf{3.8} & \textbf{4.9} & \textbf{6.9} \\
\midrule
\multirow{3}{*}{LOTUS} & \multirow{3}{*}{MMQ} & Codex (model) & 14.0 & 22.1 & 56.0 & 0.0145 & 16.5 & 1.0 & 1.5 & 3.9 \\
 &  & Codex (agent) & 15.0 & 27.7 & \textbf{73.0} & 0.2620 & 169.4 & 0.1 & 0.1 & 0.3 \\
 &  & \ours{} & \textbf{45.0} & \textbf{54.3} & 72.0 & 0.0080 & 73.2 & \textbf{5.6} & \textbf{6.8} & \textbf{9.0} \\
\midrule
\multirow{3}{*}{Nirvana} & \multirow{3}{*}{HoQ} & Codex (model) & 20.0 & 29.8 & 53.0 & 0.0099 & 11.8 & 2.0 & 3.0 & 5.4 \\
 &  & Codex (agent) & 26.0 & 36.9 & 62.0 & 0.2484 & 126.6 & 0.1 & 0.1 & 0.2 \\
 &  & \ours{} & \textbf{66.0} & \textbf{75.8} & \textbf{86.0} & 0.0117 & 79.3 & \textbf{5.6} & \textbf{6.5} & \textbf{7.4} \\
\midrule
\multirow{3}{*}{Nirvana} & \multirow{3}{*}{MQ} & Codex (model) & 5.0 & 10.8 & 20.0 & 0.0158 & 17.1 & 0.3 & 0.7 & 1.3 \\
 &  & Codex (agent) & 16.0 & 22.8 & \textbf{37.0} & 0.2759 & 165.4 & 0.1 & 0.1 & 0.1 \\
 &  & \ours{} & \textbf{27.0} & \textbf{31.2} & 35.0 & 0.0139 & 101.5 & \textbf{1.9} & \textbf{2.2} & \textbf{2.5} \\
\bottomrule
\end{tabular}}
\end{table}

\subsection{Efficiency}
\label{sec:eval:efficiency}

The efficiency results are reported in Table~\ref{tab:codex_3way} and \ref{tab:efficiency}. Table~\ref{tab:codex_3way} compares our method and two Codex methods over the same 100 sampled queries due to long running time of Codex agent. Table~\ref{tab:efficiency} reports the average latency over all three backends per dataset as supplementary to Table 1,2,3, with the 300 queries for each dataset. $t_m$, $t_e$ and $t$ are mapping, pipeline execution and total time (sum of mapping and execution).  
In addition, the two tables include quality-per-unit-cost metrics. 
Furthermore, Table~\ref{tab:efficiency} separates two regimes. Naive and Hardcoded are intentionally simple reference baselines in higher block, while the lower block (CodeTree, Codex, and
\ours{}) contains the specialized multi-agent or coding-optimized
methods which are more comparable with \ours{}. 
Within this latter group, \ours{} achieves the best
quality-per-unit-cost on every dataset, at a mapping latency
($63$--$107$\,s) that is
up to 90\% faster than CodeTree. Comparing to Codex agent (Table~\ref{tab:codex_3way}), \ours{}'s pipeline building is up to 2x faster (311.2 VS 111.3s on Palimpzest over HybridQA). These results show that \ours{} has an acceptable efficiency in building pipelines, making it practical as an agentic approach, while maintaining the highest quality and second lowest cost among all evaluated methods. The only baseline that has both better efficiency and lower cost than \ours{} is Naive, which is often unusable due to extreme low quality (like on Palimpzest and Nirvana).     

Another observation is that \ours{}'s pipelines have the fastest execution among all methods in 3 out of 5 datasets in Table~\ref{tab:efficiency}. This means the generated pipelines by \ours{} are not only high quality, but also well optimized to be highly efficient. This further shows that our method is practical and effective. 

In summary, \ours{} takes a reasonable mapping time and tiny mapping cost to gain significantly high quality and execution-optimized pipelines. Such an approach is practical to bridge semantic operator systems with automated application workflows.

\begin{table}[!htbp]
\centering
\scriptsize
\setlength{\tabcolsep}{1.5pt}
\renewcommand{\arraystretch}{0.9}
\caption{Per-pipeline efficiency on each dataset, averaged
across the three backends (Palimpzest, LOTUS, Nirvana). 
$t_m$, $t_e$, $t$ are mapping, execution, and end-to-end
latency per question (seconds). 
\textbf{EM/\$}, \textbf{F1/\$},
\textbf{LLM/\$} are quality scores per milli-USD of mapping
cost (higher is better).
\textbf{Bold} marks the best value per
(dataset, column) within the lower group only (CodeTree,
Codex (model), \ours{}); Naive and Hardcoded are reference baselines and left unbolded.}
\label{tab:efficiency}
\resizebox{\columnwidth}{!}{%
\begin{tabular}{@{}ll ccc c ccc@{}}
\toprule
\textbf{Dataset} & \textbf{Pipeline}
& \textbf{$t_m$} & \textbf{$t_e$} & \textbf{$t$}
& \textbf{\$}
& \textbf{EM/\$} & \textbf{F1/\$} & \textbf{LLM/\$} \\
\midrule
\multirow{5}{*}{TAT-QA}
 & Naive & 29.9 & 28.2 & 58.1 & 0.0043 & 4.5 & 6.1 & 9.2 \\
 & Hardcoded & 22.5 & 20.2 & 42.7 & {---} & {---} & {---} & {---} \\
\cmidrule(l){2-9}
 & CodeTree & 134.7 & 25.0 & 159.6 & 0.0217 & 1.1 & 1.6 & 2.5 \\
 & Codex (model) & \textbf{15.6} & 20.3 & \textbf{36.0} & 0.0118 & 0.9 & 1.6 & 3.8 \\
 & \textbf{\ours{}} & 92.7 & \textbf{16.2} & 109.0 & \textbf{0.0116} & \textbf{3.2} & \textbf{4.0} & \textbf{5.3} \\
\midrule
\multirow{5}{*}{HotpotQA}
 & Naive & 27.2 & 37.3 & 64.4 & 0.0035 & 7.9 & 11.2 & 15.0 \\
 & Hardcoded & 19.9 & 16.7 & 36.6 & {---} & {---} & {---} & {---} \\
\cmidrule(l){2-9}
 & CodeTree & 138.4 & 37.5 & 175.8 & 0.0243 & 1.4 & 1.8 & 2.3 \\
 & Codex (model) & \textbf{12.6} & 16.2 & \textbf{28.7} & \textbf{0.0106} & 2.0 & 3.0 & 5.1 \\
 & \textbf{\ours{}} & 78.0 & \textbf{16.1} & 94.0 & 0.0108 & \textbf{4.8} & \textbf{6.1} & \textbf{8.0} \\
\midrule
\multirow{5}{*}{HybridQA}
 & Naive & 28.3 & 59.0 & 87.3 & 0.0036 & 4.8 & 6.4 & 7.8 \\
 & Hardcoded & 26.9 & 57.6 & 84.5 & {---} & {---} & {---} & {---} \\
\cmidrule(l){2-9}
 & CodeTree & 159.7 & \textbf{53.3} & 213.1 & 0.0255 & 1.0 & 1.3 & 1.6 \\
 & Codex (model) & \textbf{18.1} & 54.9 & \textbf{73.0} & 0.0166 & 1.1 & 1.5 & 2.1 \\
 & \textbf{\ours{}} & 117.0 & 66.6 & 183.5 & \textbf{0.0156} & \textbf{2.6} & \textbf{3.2} & \textbf{3.8} \\
\midrule
\multirow{5}{*}{MMQA}
 & Naive & 27.6 & 28.6 & 56.1 & 0.0034 & 3.8 & 4.4 & 5.2 \\
 & Hardcoded & 25.9 & 31.1 & 57.0 & {---} & {---} & {---} & {---} \\
\cmidrule(l){2-9}
 & CodeTree & 166.7 & 28.4 & 195.1 & 0.0290 & 0.6 & 0.7 & 0.9 \\
 & Codex (model) & \textbf{15.2} & 19.0 & \textbf{34.3} & 0.0148 & 1.0 & 1.3 & 2.2 \\
 & \textbf{\ours{}} & 87.8 & \textbf{15.9} & 103.7 & \textbf{0.0116} & \textbf{2.9} & \textbf{3.5} & \textbf{4.3} \\
\midrule
\multirow{5}{*}{ManyModalQA}
 & Naive & 24.7 & 15.3 & 40.0 & 0.0041 & 7.9 & 9.5 & 12.5 \\
 & Hardcoded & 16.6 & 14.8 & 31.4 & {---} & {---} & {---} & {---} \\
\cmidrule(l){2-9}
 & CodeTree & 107.3 & 20.1 & 127.4 & 0.0191 & 2.1 & 2.5 & 3.3 \\
 & Codex (model) & \textbf{12.2} & \textbf{12.1} & \textbf{24.3} & 0.0109 & 2.1 & 3.1 & 5.3 \\
 & \textbf{\ours{}} & 67.2 & 12.2 & 79.4 & \textbf{0.0092} & \textbf{5.2} & \textbf{6.2} & \textbf{7.4} \\
\bottomrule
\end{tabular}}
\end{table}

\subsubsection{Efficiency of Each Phase}
We further measure the fraction of each phase's latency over the whole mapping latency in \ours{}, finding that \textbf{Phase~A is the bottleneck inside \ours{}.}
Table~\ref{tab:per-phase} reports the detailed latency fractions. Phase~A is the bottleneck
on every dataset, taking $53\%$--$61\%$ of the mapping time. Within Phase~A, the entity
grounder dominates because its input is the
largest: it sees the question plus the full data context
(table rows and previews of every text and image passage) and
must emit a structured labeling of every key phrase. Phase~B and Phase~C create less latency despite producing the most
operationally important output, because their inputs are short
structured objects (the Linker output and the action plan)
rather than data context.

\begin{table}[t]
\centering
\small
\setlength{\tabcolsep}{8pt}
\caption{Per-phase latency fraction of \ours{} on the Palimpzest
backend over selected datasets
(seconds per question). \textit{Phase~A} is the Query-Data Linker
(entity grounder $+$ optional bridge verifier), \textit{Phase~B}
is the Semantic Planner, and \textit{Phase~C} is the Code
Generator. \textbf{Bold} marks the largest latency fraction per dataset.}
\label{tab:per-phase}
\begin{tabular}{l r rrr}
\toprule
\textbf{Dataset}
& \textbf{Phase A (\%)} & \textbf{Phase B (\%)} & \textbf{Phase C (\%)} \\
\midrule
ManyModalQA & \textbf{60.6} &  15.0 & 24.4 \\
MMQA         & \textbf{53.4} & 21.3 & 25.3 \\
HybridQA     & \textbf{55.3} & 19.1 & 25.6 \\
\bottomrule
\end{tabular}
\end{table}

\paragraph{Potential Optimization on Efficiency. }
The profile in Table~\ref{tab:per-phase} suggests several
non-exclusive ways to reduce mapping latency without changing
the per-phase semantics: 
\textbf{(i)~Cheaper entity grounder.} The grounder's task is
predominantly label assignment over question spans against
data columns, closer to a structured extraction task than to
open-ended reasoning; a smaller, faster model
(e.g., \texttt{gpt-5-nano}) likely retains most of its accuracy
at a fraction of the current latency.
\textbf{(ii)~Lexical pre-filter for bridge verification.}
A cheap inverted-index or BM25 pre-filter can eliminate
candidates that lexically cannot match any data items, leaving the
LLM to confirm only the survivors that match at least one data item.
\textbf{(iii)~Prompt caching for Phase~B and Phase~C.}
Both phases share a large invariant prefix per
(backend,~dataset) pair (the planning hints and the reference
document, respectively). So prompt-caching APIs would amortize the
prefix's Time-to-first-token (TTFT) across all questions in a dataset.
In this paper, we report the un-optimized mapping latency without the solutions above so that the metrics in Table~\ref{tab:efficiency} and \ref{tab:codex_3way} reflect the
\emph{algorithm} rather than the engineering advantages.

\begin{table*}[tbp]
\centering
\small
\setlength{\tabcolsep}{3pt}
\caption{Ablation of \ours{}'s three components (PZ: Palimpzest) on MMQA and
TAT-QA across all backends. Quality metrics (EM, F1, LLM) in percentage without \%. Each row disables the listed component(s), and the
unmodified \ours{}'s numbers are reported in
Tables~\ref{tab:main_pz}--\ref{tab:main_nirvana}.
\textbf{Bold} marks the lowest score in each (backend, dataset,
metric) column.}
\label{tab:ablation}
\resizebox{\textwidth}{!}{%
\begin{tabular}{l ccc ccc ccc ccc ccc ccc}
\toprule
& \multicolumn{3}{c}{\textbf{PZ MMQA}}
& \multicolumn{3}{c}{\textbf{PZ TAT-QA}}
& \multicolumn{3}{c}{\textbf{LOTUS MMQA}}
& \multicolumn{3}{c}{\textbf{LOTUS TAT-QA}}
& \multicolumn{3}{c}{\textbf{Nirvana MMQA}}
& \multicolumn{3}{c}{\textbf{Nirvana TAT-QA}} \\
\cmidrule(lr){2-4} \cmidrule(lr){5-7} \cmidrule(lr){8-10}
\cmidrule(lr){11-13} \cmidrule(lr){14-16} \cmidrule(lr){17-19}
\textbf{Configuration}
& EM & F1 & LLM
& EM & F1 & LLM
& EM & F1 & LLM
& EM & F1 & LLM
& EM & F1 & LLM
& EM & F1 & LLM \\
\midrule
\ablate{Linker}
& 49.7 & 54.6 & 58.7
& 40.3 & 51.0 & 65.0
& 21.4 & 28.8 & 44.8
& 26.7 & 37.4 & 65.3
& 34.7 & 38.3 & 43.7
& 25.3 & 31.6 & 38.7 \\
\ablate{Plan}
& 36.3 & 45.1 & 54.7
& 32.7 & 44.3 & 58.3
& \textbf{4.3} & \textbf{9.5} & \textbf{20.3}
& \textbf{24.0} & \textbf{30.0} & \textbf{42.3}
& \textbf{8.3} & \textbf{9.3} & \textbf{10.3}
& 4.0 & 5.4 & 7.3 \\
\ablate{L\&P}
& 31.0 & 35.8 & 40.7
& 41.3 & 54.2 & 72.0
& 7.0 & 12.2 & 23.7
& 29.0 & 36.6 & 47.0
& 10.0 & 10.7 & 11.3
& \textbf{3.7} & \textbf{3.9} & \textbf{4.3} \\
\ablate{Ref}
& \textbf{19.7} & \textbf{22.3} & \textbf{25.7}
& \textbf{26.7} & \textbf{32.1} & \textbf{46.0}
& 32.1 & 38.6 & 49.5
& 37.7 & 46.2 & 60.7
& 22.7 & 27.6 & 36.0
& 27.3 & 34.4 & 46.7 \\
\bottomrule
\end{tabular}%
}
\end{table*}

\subsection{Ablation Studies}
\label{sec:eval:ablation}

We ablate the three critical building blocks of \ours{}--the Query-Data
Linker (Phase~A), the action plan (Phase~B), and the
auto-generated reference document (Phase~C input), by
disabling each component independently while holding the rest
of the workflow fixed. \textbf{\ablate{Linker}} presents the workflow that skips Phase~A
entirely, where the Planner sees only the question and the data schema.
\textbf{\ablate{Plan}} skips Phase~B, where the Code Generator instead receives the Linker's output (entity groundings and verified bridges) directly as part of the prompt. \textbf{\ablate{L\&P}}
ablates both phases A and B, and the Code Generator sees only the question and data schema. \textbf{\ablate{Ref}} replaces the auto-generated
reference document input to Phase~C with the original official backend documentation, with no
golden pipeline examples, no dataset profile, and no
backend-specific constraints, while still keeping the Phase~B action plan as part of input to Phase~C.

We report ablations on MMQA and TAT-QA across all three
backends (Table~\ref{tab:ablation}). The two datasets exercise
the most complex reasoning patterns (cross-source Compose for
MMQA, numerical aggregation with hybrid context for TAT-QA),
making them the most informative stress test. We organize the
discussion around three claims: each phase has distinct impact (Section~\ref{sec:eval:abl-independent}), the
contribution of the reference document grows with how unusual
the backend's API is (Section~\ref{sec:eval:abl-portability}), and
the Linker and Planner are coupled rather than additive
(Section~\ref{sec:eval:abl-coupling}).

\subsubsection{Each phase has distinct impact.}
\label{sec:eval:abl-independent}

Disabling a phase usually degrades quality, but the ablation
table also contains a few non-monotonic cases. The broad pattern
is nevertheless diagnostic: the drops overall align with the
role each phase plays in the \ours{}.

\emph{The Planner is overall the most consequential component.} Removing it (\ablate{Plan}) makes the
largest single-phase drop on both LOTUS and Nirvana, e.g., on LOTUS MMQA the LLM-as-judge score collapses from $51.2$ to $20.3$ ($-$30.9),
and on Nirvana MMQA from $36.0$ to $10.3$ ($-$25.7).
Palimpzest shows a smaller drop ($-$8.6 on MMQA). This
asymmetry reflects how each backend handles
under-specification: when no explicit plan is
supplied, the Code Generator must simultaneously decide
\emph{what} to compute and \emph{how} to write it. So the more the backend API requires, the higher chance that mistakes occur in the generated pipelines. 
Palimpzest's operators allow free-text predicates while LOTUS and Nirvana requires more, like one predicate has to include the column names it applies to. So missing plan has more impact on LOTUS and Nirvana than Palimpzest.   
This also indicates that the planning phase contributes information the Code Generator cannot reliably infer on its own.

%
%

\emph{The Linker helps modestly on its own, but its impact grows
when the downstream phases weaken.}
Dropping only the Linker (\ablate{Linker}) costs little on
Palimpzest ($-$4.6 on MMQA, $-$2.7 on TAT-QA): when a question
already names the entities it asks about, the downstream phases
can still locate them without the Linker. The Linker mainly helps
on harder questions whose key entity is never explicitly stated in the question and must instead be discovered in the data. The full value of the Linker becomes more visible when the downstream phases are also weakened (Section~\ref{sec:eval:abl-coupling}).

\subsubsection{The reference document matters more for less familiar backends.}
\label{sec:eval:abl-portability}

The \ablate{Ref} ablation shows that the contribution of the
auto-generated reference document is significant when a backend's
API differ from ``standard'' Python operations. On
Palimpzest, whose API style (lazy datasets, custom
schema-bound operators) is the least conventional, \ablate{Ref}
shows the largest drop on both datasets ($-$37.6 on MMQA,
$-$21.7 on TAT-QA). On LOTUS and Nirvana, whose DataFrame-style
and functional-style APIs are closer to popular programming patterns that LLM has seen during pretraining, the effect is smaller and just occasionally non-monotonic.
Therefore, the reference document bridges the gap when the LLM is not familiar with the backend syntax, and contributes less where pretraining already covers the API conventions. The reference document mechanism also contributes directly to backend portability: the same Phase~A
and Phase~B logic does not vary on backends, and only the Phase~C reference document change based on the backend. Most configuration of \ours{} keeps consistent across backends. 

\subsubsection{Linker and Planner are coupled, not additive.}
\label{sec:eval:abl-coupling}
The combined ablation \ablate{L\&P} behaves non-monotonically and
suggests an interaction between Linker and Planner. On Palimpzest TAT-QA \ablate{L\&P} has $72.0$
LLM-as-judge score, \emph{higher} than \ablate{Plan} alone ($58.3$) or \ablate{Linker} alone (65.0). The reason is when only the Plan is removed, the Linker output is input to Phase~C and occasionally biases the model toward
over-specific filters (e.g., narrow text predicates that match
no passages), while removing the Linker as well leaves the model more loosely constrained, sometimes producing simpler and robust pipelines.
This suggests Phase~A and Phase~B are \emph{coupled}: the Plan
validates and structures the Linker's evidence. Specifically, Linker output without Planner to discipline it can
hurt, and a Plan without Linker output to ground it loses much of
the informational content. 

Taken together, the four ablation conditions show that the
three phases contribute distinct benefits.
The Linker supplies entity-level evidence that cannot be easily captured by following phases. The action
plan in Phase~B structures the evidence into a backend-agnostic operator sequence. The reference
document closes the gap between the abstract plan and a
specific backend's API, and absorbs some
compilation burden depending on how unusual the API is. 
\section{Conclusion}
\label{sec:conclusion}
We presented \ours{}, one of the first middlewares enabling automated AI workflows to fully utilize semantic operator systems for query processing, by automatically compiling
natural-language questions into executable semantic operator pipelines. 
We design a three-phase workflow with related optimizations. Such design guarantees high effectiveness, low cost, and good cross-backend compatibility.   
Evaluation shows that \ours{} achieves highest quality over baselines, with tiny cost and reasonable efficiency to build semantic operator pipelines for different queries, especially complex multi-hop or multi-modal workload. We see \ours{} as a big step towards seamlessly integrating semantic operator systems into automatic AI workflow, such that the AI applications can benefit from the latest advance in semantic data systems.  

\bibliographystyle{ACM-Reference-Format}
\bibliography{references}

\end{document}